\documentclass[sigconf, nonacm]{acmart}

\usepackage{subcaption}
\usepackage{caption}
\usepackage{listings}
\usepackage{xcolor}
\usepackage{xspace}
\usepackage{todonotes}
\usepackage{tikz}
\usepackage{pgfplots}
\usepackage{hyperref}
\usepackage[capitalise]{cleveref}
\crefformat{section}{\S#2#1#3}
\crefformat{subsection}{\S#2#1#3}
\crefname{figure}{fig.}{figs.}
\Crefname{figure}{Fig.}{Figs.}
\crefname{table}{tab.}{tabs.}
\Crefname{table}{Tab.}{Tabs.}
\hypersetup{colorlinks=true, urlcolor=blue}

\definecolor{jinjapurple}{RGB}{120,45,160}
\definecolor{pyblue}{RGB}{30,80,160}

\lstdefinestyle{jinjapython}{
  language=Python,
  basicstyle=\ttfamily\scriptsize,
  columns=fullflexible,
  keepspaces=true,
  showstringspaces=false,
  breaklines=false,
  keywordstyle=\color{pyblue}\bfseries,
  stringstyle=\color{black},
  moredelim=**[s][\color{jinjapurple}\bfseries]{\{\%}{\%\}},
  moredelim=**[s][\color{jinjapurple}\bfseries]{\{\{}{\}\}},
  frame=single,
  numbers=left,
  numberstyle=\tiny,
  numbersep=5pt,
  xleftmargin=1.5em,
  framexleftmargin=1.2em,
  xleftmargin=0.06\columnwidth,
  xrightmargin=0.06\columnwidth,
  captionpos=b
}

\newcommand{\circnum}[1]{%
  \tikz[baseline=(char.base)]{
    \node[shape=circle,fill=black,text=white,inner sep=0.6pt,
    minimum size=0.8em,font=\scriptsize\bfseries] (char) {#1};
  }%
}

\newcommand\system{\textsc{Bolo}\xspace}

\newcommand{\observation}[1]{%
  \vspace{0.5mm}\noindent\underline{\textit{#1}}\enspace%
}

\pgfplotsset{compat=1.18}

\definecolor{colorBaseline}{RGB}{55,103,149}
\definecolor{colorRepairOne}{RGB}{114,188,213}
\definecolor{colorRepairTwoSOne}{RGB}{255,208,111}
\definecolor{colorRepairTwoSTwo}{RGB}{231,98,84}

\begin{document}

\title{Bolo: Verified Model Hub for Next-Generation AI Databases}
% \title{Large-Scale Agentic Synthesis of Verified Inference Pipelines}
\subtitle{Extended Abstracts}

% \author{Yunqi Li \quad Ila Petrovic \quad Yongjoo Park}
% \email{{yunqili4,ilacp2,yongjoo}@illinois.edu}
% \affiliation{%
%   \institution{University of Illinois Urbana-Champaign}
% } 

\author{Yunqi Li}
\email{yunqili4@illinois.edu}
\affiliation{%
  \institution{Univ.~of Illinois Urbana-Champaign}
}

\author{Ila Petrovic}
\email{ilacp2@illinois.edu}
\affiliation{%
  \institution{Univ.~of Illinois Urbana-Champaign}
}

\author{Yongjoo Park}
\email{yongjoo@illinois.edu}
\affiliation{%
  \institution{Univ.~of Illinois Urbana-Champaign}
}

%%
%% The "author" command and its associated commands are used to define the authors and their affiliations.

\begin{abstract}
Verified, ready-to-use inference pipelines are a cornerstone of future AI databases.
They allow multi-modal databases to incorporate specialized language, vision, and tabular models that can deliver both high accuracy and efficiency. 
Unfortunately, existing model platforms such as Hugging Face fall short of this goal. 
While they host millions of model repositories, many contain only raw weights without runnable pipelines. 
Even well-documented models often fail due to missing dependencies, unsupported model classes, or incorrect task assignments. Moreover, different models fail for different reasons, with no uniform solution. 
Constructing a large-scale, verified model hub is nearly impossible with human effort alone.

We argue that AI agents can achieve this at scale. We present \system, a model platform that hosts verified, ready-to-use inference pipelines, powered by a multi-stage agentic system for model remediation. 
For models that fail under standard usage, the agent inspects errors and repairs broken pipelines (Type~I). 
For models outside the scope of existing interfaces, it synthesizes pipelines from scratch using model metadata and
documentation (Type~II \& III). 
To prevent incorrect pipelines from entering the database, the agent applies multi-stage verification---checking not only program structure but also semantic model behavior, ensuring pipelines produce meaningful outputs rather than merely executing without error. 
In preliminary experiments, \system achieves 97.27\% and 86.08\% runnable coverage for Type~II and Type~III models, respectively, 
demonstrating that agentic synthesis with targeted verification can transform large collections of unusable model weights into a verified database of ready-to-use inference pipelines.
The preliminary database is open-sourced at \textcolor{blue}{\url{https://bolobao.ai/}}.

% \redtext{Yognjoo TBD} Public model hubs contain millions of machine learning models, but many of them cannot be used through a reliable end-to-end inference workflow. Existing interfaces such as the Transformers pipeline API provide broad support, yet many models still fail because of missing dependencies, unsupported model classes, incomplete documentation, or incorrect task assignment. In this paper, we share our recent findings that many such failures can be repaired through agentic synthesis with targeted verification. We propose a multi-stage agentic system that translates unusable model repositories into verified inference pipelines. The derived inference pipelines are then checked by runtime error verification and code hallucination verification to ensure that they not only execute, but also implement meaningful model inference. In preliminary experiments, our system improves runnable coverage for Type I models and achieves 97.27\% and 86.08\% coverage (passing runtime error verification) for Type II and Type III models, respectively. These results suggest that agentic systems can help build large-scale curated databases of usable machine learning models.
\end{abstract}

\maketitle

\section{Introduction}

Specialized models can greatly strengthen multimodal data processing.
Unlike GPT-like frontier models~\cite{gpt,claude,gemini}, specialized language/vision/tabular models offer high accuracy at a fraction of compute cost for target domains (e.g., legal, medical).
This opportunity is actively leveraged by AI-powered databases.
For example, DocETL~\cite{docetl} decomposes a natural language request into smaller subtasks, where each subtask may be processed by a more efficient model.
Data science agents (e.g., DS-Agent~\cite{ds-agent}, AutoKaggle~\cite{autokaggle}, notebook agent~\cite{kishu-agent}) can dynamically synthesize model-based scripts, which may include pre-trained public models.
These multimodal databases will become more successful if a variety of specialized models are readily available for user-specific tasks.

Unfortunately, \emph{\textbf{existing model platforms fail to offer readily available models}}.
Platforms like Hugging Face~\cite{huggingface_hub} and ModelScope~\cite{modelscope} host millions of models, aiming to provide end-to-end services such as fine-tuning, inference, and deployment.
Nearly any researcher can create model \emph{repositories} and upload model (weights) to share them publicly.
While the resources are extremely valuable, many repositories only contain \emph{weights} rather than runnable inference pipelines.
As we analyze in \cref{sec:motivation_scope}, a majority of the models are not runnable due to architectural mismatch, missing configuration files, poor documentation, and so on.
Moreover, manually fixing the issues is non-trivial because different models fail due to different reasons, requiring model-specific repairs.
% As the vast number of models have heterogeneous architectures, and the critical reference for utilizing models are voluntarily maintained by users (e.g., \verb|README.md|). Consequently, many of the models do not support seamlessly end-to-end usage. 
% In order to unlock their potential, we need a large-scale, curated database of high-quality models that users can easily download, compare, and deploy.

% Nowadays, machine learning models are widely applied across various fields. 

\paragraph{Our Goal}
To address this, we are building \emph{\textbf{a new model platform (called \system) that hosts verified, ready-to-use models}}.
Our core idea is as follows: while it is prohibitively expensive for humans to test and repair millions of (broken) models, AI agents~\cite{zhong2026actionenginereactiveprogrammaticgui, russo2025deepresearchnewanalytics, chen2026stratusmultiagentautonomousreliability, zeng2024simplefastwayhandle, yuan2026transagentenhancingllmbasedcode, xia2025livesweagentsoftwareengineeringagents} could achieve it in a scalable way.
That is, if we could build a novel AI agent specialized for model remediation (e.g., inspecting errors, generating model-specific code, and verifying end-to-end inference), we could use the agent to turn millions of (broken) models into verified, ready-to-use inference pipelines.
These verified models can then be directly plugged into AI databases to offer compelling accuracy/efficiency for a wide range of multimodal tasks.

\paragraph{Challenge}
Our project faces the following challenges (\textbf{C1--C3}). \textbf{(C1)} Existing model interfaces are incomplete. Interfaces like Hugging Face's \verb|Transformers|~\cite{wolf2020huggingfacestransformersstateoftheartnatural} cover only a subset of hosted models---only 3,163 of 5,293 object detection models are supported---and extending such manually curated collections to millions of continuously uploaded models is infeasible. 
\textbf{(C2)} Even well-documented models have broken or missing inference pipelines. Reference usages found in documentation and model cards are frequently outdated, incomplete, or scattered, causing failures due to missing dependencies, unsupported model classes, or incorrect task assignments. 
\textbf{(C3)} Verifying a generated pipeline requires unique semantic checks. 
Conventional software engineering may only involve deterministic tests.
Yet, ML models are black boxes. A pipeline may be syntactically correct and may run without error; however, it could produce semantically meaningless outputs.
% A reliable system must verify not only that generated code executes, but that it performs intended inference.

\paragraph{Our Approach}
To achieve a large-scale hub of verified models, we have built (and are continuously improving) a multi-stage agentic system that directly addresses each challenge above.
\textbf{(A1)} To overcome the coverage limitation of \verb|Transformers|, our system synthesizes model-specific inference pipelines from scratch for models outside its supported scope, using each model's metadata and documentation as context.
\textbf{(A2)} For models within the \verb|Transformers| scope that fail under standard usage, the system inspects errors and repairs their pipelines by resolving missing dependencies, correcting model class assignments, and updating outdated usage patterns.
\textbf{(A3)} To prevent incorrect pipelines, we introduce multi-stage verification procedures that go beyond structural checks: the system tests model behavior end-to-end, ensuring that pipelines not only execute but also produce semantically meaningful outputs under the intended task setting.

\begin{figure*}[t]
    \centering
    \begin{subfigure}[t]{0.26\textwidth}
        \centering
        \includegraphics[width=\linewidth]{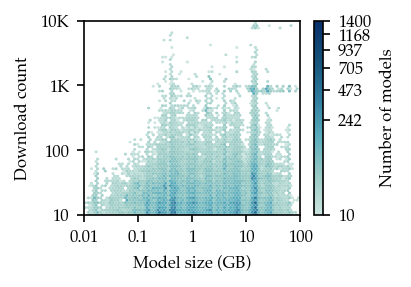}
        \caption{Model size vs. usage}
        \label{fig:motivation:observation:model_size_usage}
    \end{subfigure}
    \hfill
    \begin{subfigure}[t]{0.35\textwidth}
        \centering
        \includegraphics[width=\linewidth]{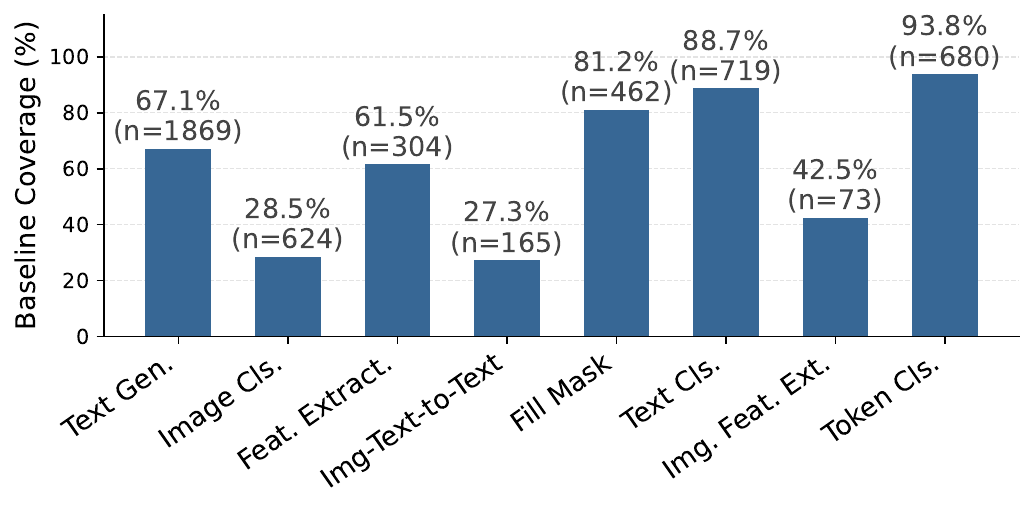}
        \caption{Fraction of runnable models in baseline}
        \label{fig:motivation:observation:baseline_coverage}
    \end{subfigure}
    \hfill
    \begin{subfigure}[t]{0.36\textwidth}
        \centering
        \includegraphics[width=\linewidth]{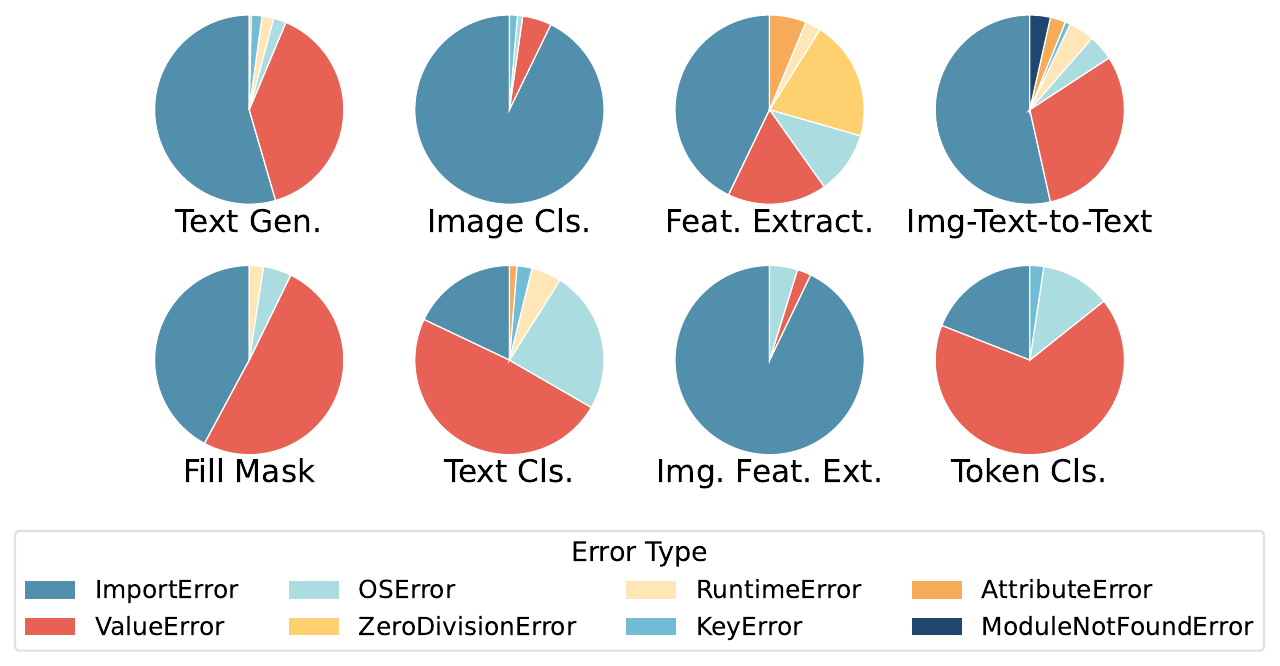}
        \caption{Failed models - error type distribution}
        \label{fig:motivation:observation:error_dist}
    \end{subfigure}
    \caption{We analyzed Hugging Face models to understand (a) the relationship between model sizes and their usages, (b) the fractions of runnable and non-runnable models, and (c) the distribution of unique error types. The volume of models to support and the diversity of errors are vast, thus motivating agentic approaches toward a verified model hub.}
    \label{fig:motivation:observation}
\end{figure*}

\section{Motivation \& Scope} \label{sec:motivation_scope}

We understand why fixing broken models with humans is nearly impossible (\cref{sec:motivation_scope:motivation}).
Then, we clarify the scope of this work (\cref{sec:motivation_scope:scope}).

\subsection{Motivation} \label{sec:motivation_scope:motivation}

To understand the current state, we collected 2,250,578 models from Hugging Face.\footnote{Models were collected as of February 6, 2026.} We make three observations (O1-3).

\observation{O1: Both small/large models are widely used.}
\Cref{fig:motivation:observation:model_size_usage} illustrates the relationship between model size and usage. First, models across a broad range of sizes, and both small and large models are widely used. Second, a substantial fraction of models receive relatively low usage, with download counts between 10 and 1K.

\observation{O2: Many models are unusable.}
\verb|Transformers|~\cite{wolf2020huggingfacestransformersstateoftheartnatural} covers 810,572 models, and the \verb|pipeline| API aims to provide a universal interface for all \verb|Transformers| models. To set up it as a baseline, we selected models with downloads exceeding the average (which, after excluding models with zero downloads, stood at 48,754). Then, we copied the \verb|pipeline| inference code with dependencies from the official documentations of Hugging Face,\footnote{\url{https://huggingface.co/docs/transformers/en/tasks/<TASK_NAME>}} to run inference on GPU. 
\Cref{fig:motivation:observation:baseline_coverage} shows the fraction of runnable models (i.e., coverage) for each task. Many models failed due to various errors (see below). 

\observation{O3: No common fix; we need both agents \& human experts.}
We collected those non-runnable models and analyzed the errors. For ease of presentation, we report only error types that occur more than three times across all tasks (without error details). \Cref{fig:motivation:observation:error_dist} shows that 1) all tasks have a substantial number of \verb|ImportError| cases, indicating missing task-specific dependencies. 2) Each task still exhibits a diverse set of error types, and different tasks show distinct error distributions. For example, the most frequent error type for image classification models is \verb|ImportError|, whereas for token classification models it is \verb|ValueError|. For the diverse error types, human is good at identifying and configuring missing dependencies while agents are suitable to understand other code-related errors then patch the current \verb|pipeline| API.

\subsection{Scope of This Work} \label{sec:motivation_scope:scope}

% Define Type I, II, & III models and illustrate the sigfinicance
\paragraph{Supported Models}

The 2.8M models on Hugging Face can be grouped into three categories. \textbf{Type I} models (studied in \cref{sec:motivation_scope:motivation}) are directly supported by the \verb|pipeline| API and span 22 tasks; the task associated with each model is indicated by the \verb|pipeline_tag| field in its model card. \textbf{Type II} models are still tagged as \verb|Transformers|, but either lack a clear \verb|pipeline_tag| or correspond to tasks outside the current scope of \verb|pipeline| support. \textbf{Type III} models are non-\verb|Transformers| models. For (a small number of) large models, engines like vLLM~\cite{kwon2023efficient} provide strong support. We therefore exclude them from our scope.

% Clarify the difference between transformers & ours
\paragraph{Verifying inference pipelines}

When executing agent-generated inference programs, different types of errors may arise. The first type is the \textbf{runtime error}, which is raised during program execution and indicates that the generated code fails to execute correctly. 
The second type is \textbf{code hallucination}~\cite{Tian_Yan_Yang_Zhao_Chen_Wang_Luo_Ma_Song_2025, kahati2026detecting}: even if the program runs without runtime errors, it may fail to implement the intended inference semantics. 
For example, a pipeline generated for model A may incorrectly use model B as the model instance during inference. 
In addition, a correct inference pipeline does not guarantee correct model outputs, since the model itself may be buggy or misaligned with its intended task. 
We therefore introduce \textbf{test-specific data verification}, where task-specific test cases are prepared for each fine-grained task. 
The resulting outputs are evaluated by human experts or an LLM-as-judge against the expected results. We currently focus on the first two verifications.

\section{Agentic System} \label{sec:agentic_system}

\begin{figure*}[t]
    \centering
    \begin{subfigure}[t]{0.645\textwidth}
        \centering
        \includegraphics[width=\linewidth]{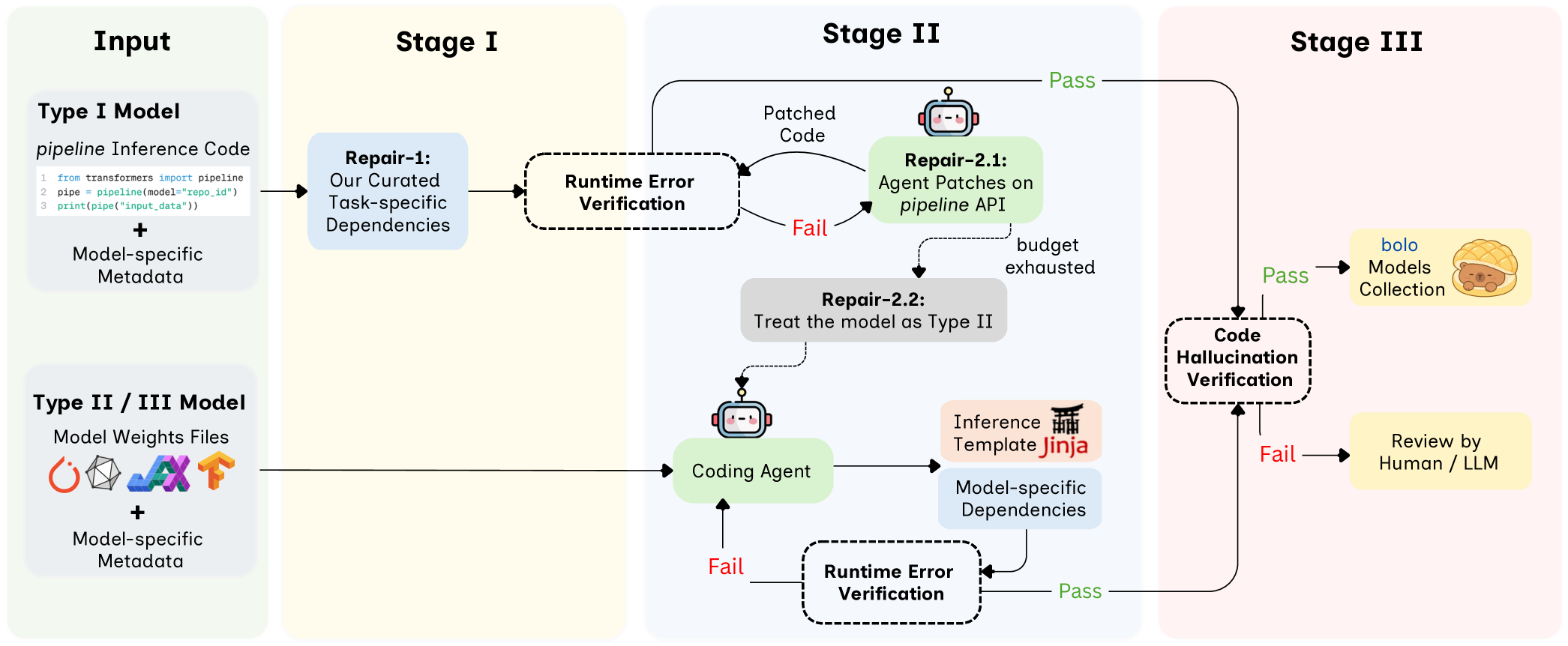}
        \caption{Workflow of our agentic system}
        \label{fig:agentic_system:workflow}
    \end{subfigure}
    \hfill
    \begin{subfigure}[t]{0.345\textwidth}
        \centering
        \includegraphics[width=\linewidth]{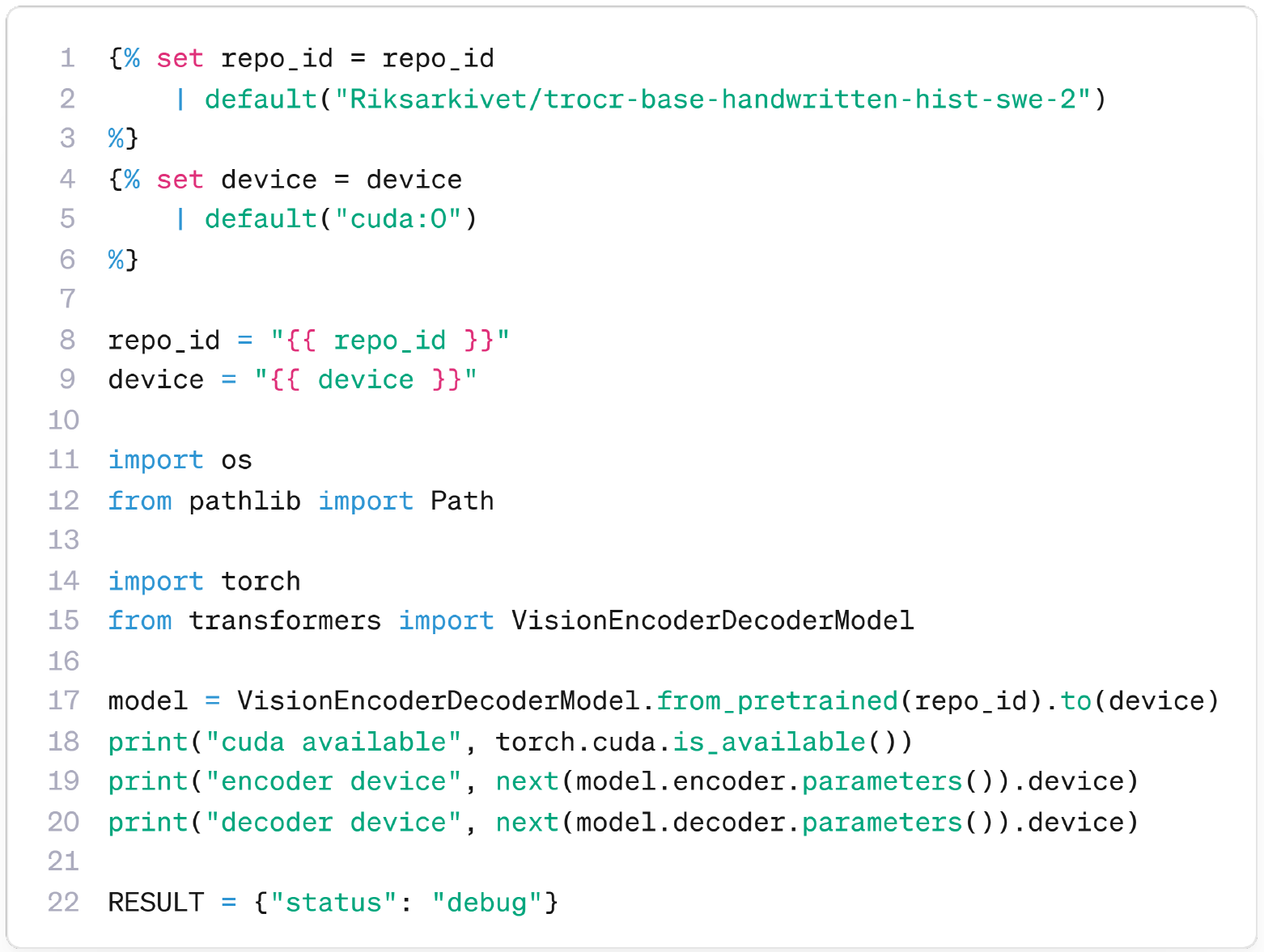}
        \caption{Hallucinated inference template}
        \label{fig:agentic_system:code}
    \end{subfigure}
    \caption{Our agentic system converts unusable model-weight repositories into verified inference pipelines. As shown in (a), Type I models are repaired by patching the pipeline API, while Type II and Type III models are handled by generating model-specific Jinja2 inference templates. The generated code is then passed to a hallucination-detection stage, where hallucinated templates, such as the example in (b), are filtered out and sent for review.}
    \label{fig:combined}
\end{figure*}

To construct a large-scale verified model hub, we introduce a specialized agentic system.
\Cref{fig:agentic_system:workflow} presents its overall workflow.
Our system adopts a three-stage process to compose the inference pipeline for a given model. 
The system takes different inputs depending on the model type. 
For Type~I models, our system only considers models that fail in the baseline described in \cref{sec:motivation_scope:motivation}. 
It starts from the original \verb|pipeline|-based inference code used in the baseline, together with model-specific metadata such as the \verb|README.md| file. 
For Type II \& III models, the system instead takes model weight files and model-specific metadata as input. 

The first (\cref{sec:agentic_system:stageI}) and second (\cref{sec:agentic_system:stageII_I}, \cref{sec:agentic_system:stageII_II}) stages focus on inference pipeline generation and runtime error verification. 
The generation strategy differs across model types: 
for Type I models, the system \textbf{patches} the existing \verb|pipeline|-based code; 
for Type II \& III models, it \textbf{generates} inference pipelines as Jinja2 templates along with model-specific dependencies. 
Finally, Stage III (\cref{sec:agentic_system:stage3}) performs code hallucination verification on all inference programs from Stage II. 

\subsection{Stage I: Repair Dependencies} \label{sec:agentic_system:stageI}

As shown in \cref{sec:motivation_scope:motivation}-O3, each model may require additional model-specific dependencies for inference (e.g., a \textit{text generation} model may rely on a tokenizer that requires an additional Python package). 
Indeed, identifying model-specific dependencies could repair a substantial portion of previously non-runnable Type I models.

\paragraph{Repair-1.}

For a non-runnable Type I model, our system uses the baseline \verb|pipeline| inference program as input. It identifies the model's task and runs the program in a fresh virtual environment with pre-configured task-specific dependencies. These dependencies are prepared during baseline verification by manually analyzing task-level error messages and identifying missing packages. The program is then executed once as runtime error verification. If it succeeds, it proceeds to Stage III for hallucination detection, which should pass since the program remains unchanged. Otherwise, the program, dependencies, and model-specific metadata are sent to Stage II for further repair.

% Given a non-runnable Type I model, our system only takes its \verb|pipeline| inference program from baseline as input, then identifies the model's corresponding task and loads pre-configured task-specific dependencies (i.e., \verb|requirements.txt|) in a fresh virtual environment (venv). Each \verb|requirements.txt| is configured during the baseline verification, a human expert will collect error messages of all models of the task, analyze and identify the missing dependencies. Then, the system will execute the inference program in configured venv with specific input (e.g., an image for \textit{object detection} models) as a one-pass runtime error verification. Upon success, the system will enter Stage III to detect code hallucination on the inference program (it will always pass since we haven't made any changes on the program yet). Otherwise, the inference program, configured dependencies and model-specific metadata will be sent to Stage II for further fixing.

\subsection{Stage II: Finalize Type I Model Repair} \label{sec:agentic_system:stageII_I} 

In Stage II, our system invokes agents to derive runnable inference programs for Type I models in two steps.

\paragraph{Repair-2.1.}

For a Type I model failing at Stage I runtime error verification, our system invokes an agent to patch the inference program (e.g., editing the model class definition to add missing configuration variables), and keep the inference library and dependencies unchanged. The tool set of our patching agent involves runtime error verification and read-only file system functions to explore the metadata. A budget on the number of tool calls is applied to the patching agent; once the budget is exceeded, the model along with the current state will be sent to Repair-2.2.

\paragraph{Repair-2.2.}
For Type I models failing at Repair-2.1 runtime error verification, our system will treat it as a Type II model, and send only the model's metadata and weights files to the generation workflow (see below).

\subsection{Stage II: Repair Type II \& III Models} \label{sec:agentic_system:stageII_II} 

Given a Type II or III model, our system takes its weights files and metadata as input, then invokes a coding agent. The tool set of this agent includes runtime error verification and venv management functions (e.g., install/uninstall packages). The coding agent emits the Jinja2 template along with a model-specific \verb|requirements.txt|. At the time the agent invokes runtime error verification, our system will first configure dependencies in a fresh venv then executes the template by rendering its variables (e.g., input) with default values. The same budget is applied on coding agent. Once the agent fails to derive the inference program that passes runtime error verification for a model, the system will mark the model as failed.

\begin{figure*}[t]
\centering

% Left: figure
\begin{minipage}[t]{0.66\textwidth}
\centering
\vspace{0pt}
\begin{tikzpicture}
\begin{axis}[
    ybar stacked,
    width=\linewidth,
    height=4.2cm,
    bar width=7.5pt,
    ylabel={Runnable Coverage (\%)},
    ylabel style={font=\scriptsize},
    ymin=0, ymax=112,
    xtick={0,1,2,3,4,5,6,7,8,9,10,11,12,13,14,15,16,17,18,19,20,21},
    xticklabels={
        Mask (n=14), ImgCls (n=624), ImgFeat (n=73), ImgSeg (n=69), Img2Txt (n=164), VidCls (n=26),
        Any2Any (n=9), Txt2Aud (n=14), TableQA (n=14), ZSObj (n=17), FeatExt (n=304), ZSImg (n=74),
        TextGen (n=1752), DocQA (n=4), ObjDet (n=45), FillMask (n=462), ZSCls (n=50), Depth (n=22),
        TextCls (n=719), ASR (n=264), TokCls (n=680), AudCls (n=44)
    },
    xticklabel style={
        rotate=60,
        anchor=east,
        font=\tiny,
        align=right
    },
    yticklabel style={font=\tiny},
    tick style={line width=0.3pt},
    ymajorgrids=true,
    grid style={dashed, line width=0.3pt, draw opacity=0.35},
    axis x line*=bottom,
    axis y line*=left,
    legend style={
        at={(0.5,1.04)},
        anchor=south,
        font=\tiny,
        fill=white,
        fill opacity=0.9,
        draw={gray!40},
        text opacity=1,
        cells={anchor=west},
        column sep=0.5em,
    },
    legend columns=4,
    clip=false,
    enlarge x limits=0.03,
]
\addplot[fill=colorBaseline, draw=none] coordinates {(0,0.0000) (1,28.5256) (2,42.4658) (3,47.8261) (4,27.4390) (5,50.0000) (6,0.0000) (7,50.0000) (8,64.2857) (9,58.8235) (10,61.5132) (11,72.9730) (12,71.6324) (13,75.0000) (14,77.7778) (15,81.1688) (16,82.0000) (17,81.8182) (18,88.7344) (19,92.8030) (20,93.8235) (21,93.1818)}; \addlegendentry{Baseline}
\addplot[fill=colorRepairOne, draw=none] coordinates {(0,100.0000) (1,64.7436) (2,54.7945) (3,43.4783) (4,25.0000) (5,0.0000) (6,11.1111) (7,0.0000) (8,0.0000) (9,35.2941) (10,3.6184) (11,12.1622) (12,16.8950) (13,0.0000) (14,2.2222) (15,10.8225) (16,8.0000) (17,4.5455) (18,4.8679) (19,0.3788) (20,2.0588) (21,0.0000)}; \addlegendentry{After Repair-1}
\addplot[fill=colorRepairTwoSOne, draw=none] coordinates {(0,0.0000) (1,5.9295) (2,2.7397) (3,4.3478) (4,17.0732) (5,42.3077) (6,22.2222) (7,42.8571) (8,35.7143) (9,0.0000) (10,23.0263) (11,12.1622) (12,7.1918) (13,25.0000) (14,15.5556) (15,7.3593) (16,10.0000) (17,9.0909) (18,4.8679) (19,3.7879) (20,3.2353) (21,2.2727)}; \addlegendentry{After Repair-2.1}
\addplot[fill=colorRepairTwoSTwo, draw=none] coordinates {(0,0.0000) (1,0.1603) (2,0.0000) (3,0.0000) (4,4.8780) (5,3.8462) (6,11.1111) (7,0.0000) (8,0.0000) (9,0.0000) (10,5.9211) (11,1.3514) (12,1.3128) (13,0.0000) (14,4.4444) (15,0.2165) (16,0.0000) (17,0.0000) (18,0.5563) (19,1.5152) (20,0.1471) (21,0.0000)}; \addlegendentry{After Repair-2.2}
    \node[font=\tiny, align=center, text=black!65, anchor=south] at (axis cs:0,101.20) {100.0\%};
    \node[font=\tiny, align=center, text=black!65, anchor=south] at (axis cs:1,100.56) {99.4\%};
    \node[font=\tiny, align=center, text=black!65, anchor=south] at (axis cs:2,101.20) {100.0\%};
    \node[font=\tiny, align=center, text=black!65, anchor=south] at (axis cs:3,96.85) {95.7\%};
    \node[font=\tiny, align=center, text=black!65, anchor=south] at (axis cs:4,75.59) {74.4\%};
    \node[font=\tiny, align=center, text=black!65, anchor=south] at (axis cs:5,97.35) {96.2\%};
    \node[font=\tiny, align=center, text=black!65, anchor=south] at (axis cs:6,45.64) {44.4\%};
    \node[font=\tiny, align=center, text=black!65, anchor=south] at (axis cs:7,94.06) {92.9\%};
    \node[font=\tiny, align=center, text=black!65, anchor=south] at (axis cs:8,101.20) {100.0\%};
    \node[font=\tiny, align=center, text=black!65, anchor=south] at (axis cs:9,95.32) {94.1\%};
    \node[font=\tiny, align=center, text=black!65, anchor=south] at (axis cs:10,95.28) {94.1\%};
    \node[font=\tiny, align=center, text=black!65, anchor=south] at (axis cs:11,99.85) {98.6\%};
    \node[font=\tiny, align=center, text=black!65, anchor=south] at (axis cs:12,98.23) {97.0\%};
    \node[font=\tiny, align=center, text=black!65, anchor=south] at (axis cs:13,101.20) {100.0\%};
    \node[font=\tiny, align=center, text=black!65, anchor=south] at (axis cs:14,101.20) {100.0\%};
    \node[font=\tiny, align=center, text=black!65, anchor=south] at (axis cs:15,100.77) {99.6\%};
    \node[font=\tiny, align=center, text=black!65, anchor=south] at (axis cs:16,101.20) {100.0\%};
    \node[font=\tiny, align=center, text=black!65, anchor=south] at (axis cs:17,96.65) {95.5\%};
    \node[font=\tiny, align=center, text=black!65, anchor=south] at (axis cs:18,100.23) {99.0\%};
    \node[font=\tiny, align=center, text=black!65, anchor=south] at (axis cs:19,99.68) {98.5\%};
    \node[font=\tiny, align=center, text=black!65, anchor=south] at (axis cs:20,100.46) {99.3\%};
    \node[font=\tiny, align=center, text=black!65, anchor=south] at (axis cs:21,96.65) {95.5\%};
\end{axis}
\end{tikzpicture}

\vspace{-1.2em}
\captionof{figure}{Stage-by-stage runnable coverage improvement for Type I models.}
\label{fig:exp:type_i_coverage}
\end{minipage}
\hfill
% Right: two stacked tables
\begin{minipage}[t]{0.32\textwidth}
\centering
\vspace{10pt}

\captionof{table}{Coverage and cost: Type II \& III.}
\label{tab:exp:type_ii_iii}

\vspace{-0.8em}
\resizebox{\linewidth}{!}{
\begin{tabular}{lcccc}
\toprule
& \multicolumn{2}{c}{mini-swe} & \multicolumn{2}{c}{Ours} \\
\cmidrule(lr){2-3} \cmidrule(lr){4-5}
Type & Coverage & Cost & Coverage & Cost \\
\midrule
Type II  & 93.05\% & \$36.65 & 97.27\% & \$37.26 \\
Type III & 73.12\% & \$53.60 & 86.08\% & \$134.54 \\
\bottomrule
\end{tabular}
}

\vspace{2.0em}

\captionof{table}{Code hallucination verification.}
\label{tab:exp:hallu_ver}

\vspace{-0.8em}
\resizebox{\linewidth}{!}{
\begin{tabular}{lcccccc}
\toprule
Category & TP & FP & TN & FN & Recall & Precision \\
\midrule
Type I models  & 67 & 24 (\textit{7}) & 90  & 1 & 98.5\% & 79.8\%   \\
Type II models & 71 & 48 (\textit{29}) & 119 & 0 & 100.0\% & 78.9\%  \\
\bottomrule
\end{tabular}
}

\vspace{0.5em}

\end{minipage}

\end{figure*}

\subsection{Stage III: Code Hallucination Verification} \label{sec:agentic_system:stage3}

Both patching and coding agents may generate hallucinated code that does not implement the correct inference semantics but passes the verification. A hallucinated template (shown in \cref{fig:agentic_system:code}) may bypass the input data and contain only a few meaningless \verb|print| calls. Code hallucination verification is therefore designed to filter out such templates. The filtered inference programs are then reviewed by an LLM judge or a human expert to confirm whether the filtering decision is correct. We start with a framework for characterizing non-hallucinated inference programs, and then derive the verification algorithm \textsf{HalluVer}.

\paragraph{Formal Framework}
We formalize inference semantics as a tuple $(\mathcal{X},\mathcal{Y},M)$, where $\mathcal{X}$ is the input space, $\mathcal{Y}$ is the output space, and $M$ is the model specification. To realize such semantics, a non-hallucinated inference program $P$ should contain three steps: \circnum{1} load the target model; \circnum{2} perform computation on task input; and \circnum{3} construct the final output. Let $L(P)$ denote the inference library (e.g., \verb|Transformers|) imported by $P$. And $g_P$ is $P$'s data-flow graph, where $u\rightsquigarrow_P v$ means that $u$ can flow to $v$. $P$ should satisfy the following:

{\small
\setlength{\jot}{2pt}
\[
\begin{aligned}
\exists x,m,z,y:\;& x \in \mathcal{X} \\
& {}\land m \leftarrow f(M), \quad f \in L(P)
    && \text{\circnum{1} load target model} \\
& {}\land z \leftarrow q(m,x'), \quad q \in L(P)
    && \text{\circnum{2} compute on input} \\
& {}\land y \in \mathcal{Y} \land z \rightsquigarrow_P y
    && \text{\circnum{3} construct output}.
\end{aligned}
\]
}

\noindent
Here, \(x'\) denotes the possibly pre-processed input, and the reachability condition \(z\rightsquigarrow_P y\) allows arbitrary post-processing between the raw inference output and the final result. $f$ and $q$ are model loading and computation API calls from $L(P)$.

% \noindent\underline{\textit{Algorithm for Type I models.}}
% For Type I models, the model construction and inference events should be realized through the \verb|pipeline| API. We therefore verify \(P\) by checking whether the following pattern appears in \(g_P\):

% {\small
% \setlength{\jot}{2pt}
% \[
% \begin{array}{rcl}
% \Ver_{\mathrm{I}}(P,t)=1
% &\iff&
% \exists x,v,y\in V_P:
% \Phi_x \land \Phi_v \land \Phi_y, \\[2pt]
% \Phi_x &\equiv& \Pipe(x),\\
% \Phi_v &\equiv& \Call(v,x),\\
% \Phi_y &\equiv& \RD(y)\land v\rightsquigarrow_P y .
% \end{array}
% \]
% }

% Here, \(\operatorname{Pipe}(x)\) means that \(x\) is constructed by a \verb|pipeline| call, \(\operatorname{Invoke}(v,x)\) means that \(v\) is produced by invoking \(x\), and \(\operatorname{ResultDict}(y)\) means that \(y\) is the constructed \verb|RESULT| dictionary. The reachability condition allows arbitrary post-processing between the raw inference output and \verb|RESULT|.

\paragraph{Algorithm: HalluVer}

\textsf{HalluVer} takes $P$ as input, parses $P$ into an abstract syntax tree (AST), and constructs its data-flow graph $g_P$. It then verifies whether $P$ implements the above three steps correctly. 
% First, it verifies that the program defines a valid task input $x$. Second, it searches for $f \in L(P)$ that loads the correct $M$ and binds the result to a model object $m$. Third, it searches for $q \in L(P)$ that applies the loaded model to the input $x'$, producing an intermediate result $z$. Finally, it checks whether the final output $y$ is constructed from $z$ by testing the reachability condition $z \rightsquigarrow_P y$ in $g_P$. 
If any of these checks fails, \textsf{HalluVer} flags $P$ as \texttt{hallucinated}; otherwise, $P$ passes code hallucination verification. To identify $f$ and $q$ in $L(P)$, \textsf{HalluVer} relies on an offline curated API collection $C$ for each inference library. 

% For Type II and III models, inference programs may use model-specific APIs. We instantiate the framework with syntactic and data-flow checks. Let \(L(P)\) be the imported library symbols and \(g_P\) be the partial data-flow graph of \(P\). We verify \(P\) as follows:

% % Verification for Type II & III
% {\small
% \setlength{\jot}{2pt}
% \[
% \begin{aligned}
% \Ver_{\mathrm{II/III}}(P,t)=1
% & \iff \exists A,f,q,m,z,y:\;
% \Phi_{\mathrm{in}}
% \land \Phi_{\mathrm{model}}
% \land \Phi_{\mathrm{infer}}
% \land \Phi_{\mathrm{out}},\\
% \Phi_{\mathrm{in}}
% &\equiv\;
% A\subseteq\Assign(P)\land |A|\ge4 \land H_J(A) \land H_O(A)\\
% & \land H_D(A),\\
% \Phi_{\mathrm{model}}
% &\equiv\;
% f\in L(P)\land \Load(f) \land m\leftarrow f(M) \\
% &\land\Assoc(m,M),\\
% \Phi_{\mathrm{infer}}
% &\equiv\;
% q\in L(P)\land \Infer(q)\land z\leftarrow q(\cdots),\\
% \Phi_{\mathrm{out}}
% &\equiv\;
% \RD(y)\land z\flow y .
% \end{aligned}
% \]
% }

% Here, \(H_J\), \(H_O\), and \(H_D\) denote the checks for Jinja variables, output directory, and input data, respectively. In the current implementation, \(\Phi_{\mathrm{infer}}\) only checks the presence of an inference API call from \(L(P)\), without further validating its full data flow or argument consistency.

\section{Preliminary Results} \label{sec:exp}

\subsection{Evaluation Setup} \label{sec:exp:setup}

\paragraph{Models}
% \noindent\underline{\textit{Models.}}\enspace
Following the baseline setup in \cref{sec:motivation_scope:motivation}, we collected Hugging Face models with greater-than-average download counts. Then, we excluded the models larger than 16GB and those with external service agreements.
Finally, we obtained 5,444 Type I models, 1,353 Type II models, and 1,581 Type III models. 

\paragraph{Agents}
% \noindent\underline{\textit{Agents.}}\enspace
For Type II \& III models, we compare our agentic system (\cref{sec:agentic_system:stageII_II}, noted \textsf{Ours})  to 
\textsf{mini-swe}-agent~\cite{yang2024sweagentagentcomputerinterfacesenable}. 
\textsf{mini-swe} is configured to use Apptainer sandbox for arbitrary \verb|bash| commands.
The budget for all agents is 50 tool calls. We set 20-min timeout for each model in Stage II. 
The LLM backend is \textsf{ChatGPT-5.1-CodeX-mini}.

\paragraph{Configurations.}
All experiments were conducted on an NVIDIA A40 GPU with CUDA version 12.8. 
The GPU was used for runtime error verification. LLM calls were remote.
For the baseline of Type I models, we used v5.3.0 \verb|Transformers| by Hugging Face. 

\subsection{For Some Models, Agent Boosts Coverage} \label{sec:results}

\Cref{fig:exp:type_i_coverage} presents the coverage (i.e., fraction of runnable inference programs after runtime error verification) improvements of Type I models, which echoes \Cref{fig:motivation:observation:error_dist}. The average coverage of the baseline across all tasks is 69.3\%, then after Stage I fixing (dependencies configuration) it comes to 88\%, leaving the last 10\% to be covered. Our proposed agentic strategy could resolve this ``last-mile'' problem, increasing the coverage to 96.1\% (Repair-2.1) and 97.3\% (Repair-2.2) on average.

% \begin{figure}[htbp]
%     \centering
%     \includegraphics[width=\linewidth]{figures/typeI_coverage_stages.pdf}
%     \caption{Coverage for Type I models}
%     \label{fig:exp:typeI}
% \end{figure}

\subsection{For Other Models, Agent is Essential}

For Type II \& III models, \cref{tab:exp:type_ii_iii} reports (1) the coverage after runtime error verification and (2) the total API costs in US Dollars. \textsf{Ours} reached higher coverage than \textsf{mini-swe} for both Type II \& III models. We attribute this improvement to our more task-oriented tool set used. 
That is, compared with unrestricted \verb|bash| commands, our tool design guided the agent toward inference-pipeline generation and reduced cases where the tool-call budget or wall-time limit was exhausted.

However, \textsf{Ours} incurred higher API costs than \textsf{mini-swe} on Type III models. Although our agent used fewer steps, its prompt was longer than that of \textsf{mini-swe}. Moreover, since the two agents followed the same core reasoning and action steps, the longer prompt led \textsf{Ours} to consume more reasoning tokens overall. Further investigation and cost reduction are left as ongoing work. 

% \begin{table}[htbp]
% \centering
% \caption{Coverage and cost of Type II and Type III models under mini-swe and ReAct.}
% \label{tab:type_ii_iii_coverage}
% \small
% \begin{tabular}{@{}lcccc@{}}
% \toprule
% & \multicolumn{2}{c}{\textsf{mini-swe}} 
% & \multicolumn{2}{c}{\textsf{Ours}} \\
% \cmidrule(lr){2-3} \cmidrule(lr){4-5}
% \textbf{Type} 
% & \textbf{Coverage} & \textbf{Cost} 
% & \textbf{Coverage} & \textbf{Cost} \\
% \midrule
% Type II  & 93.05\% & \$36.65  & 97.27\% & \$37.26  \\
% Type III & 73.12\% & \$53.60  & 86.08\% & \$134.54 \\
% \bottomrule
% \end{tabular}
% \end{table}

\subsection{Hallucination Verification is Important}

Finally, we evaluated \textsf{HalluVer} for verifying code hallucination in runtime-error-free inference programs for Type I and Type II models. 
We first curated an API collection $C$ for \verb|Transformers|, covering model-loading APIs and inference-related computation APIs. 
Using this collection, \textsf{HalluVer} flagged 91 hallucinated programs among 481 runnable Type I inference programs, and 119 hallucinated programs among 1,316 runnable Type II inference programs. 
After filtering these programs, the average runnable coverage was reduced to 95.3\% after Repair-2.1 and 95.6\% after Repair-2.2 for Type I models, and to 88.5\% for Type II models.

To evaluate these filtering decisions, a human expert annotated all programs flagged as \texttt{hallucinated}, together with an equally sized sample of unflagged programs. 
\Cref{tab:exp:hallu_ver} reports the resulting false positives and false negatives. 
The results show that \textsf{HalluVer} minimizes false negatives by aggressively flagging suspicious pipelines, but incurs non-negligible false positives. Specifically, 24 Type I and 48 Type II valid programs were incorrectly flagged, motivating a final human or LLM-based review before exclusion. 
The bracketed numbers denote programs outside the current \verb|Transformers|-based coverage of \textsf{HalluVer}, which we exclude when calculating precision. 

% \begin{table}[htbp]
% \centering
% \caption{Code hallucination verification.}
% \label{tab:verification_results}
% \begin{tabular}{lrrrrrr}
% \toprule
% \textbf{Category} & \textbf{TP} & \textbf{FP} & \textbf{TN} & \textbf{FN} 
% & \textbf{Recall} & \textbf{Precision} \\
% \midrule
% Type I models & 67 & 24 & 90 & 1 & 79.8\%* & 98.5\% \\
% Type II models & 71 & 48  & 119  & 0 & 81.0\%* & 100.0\% \\
% \bottomrule
% \end{tabular}
% \end{table}

\clearpage

\bibliographystyle{ACM-Reference-Format}
\bibliography{ref}

\end{document}